\documentclass[twocolumn,showpacs,preprintnumbers,amsmath,amssymb]{revtex4}
\input epsf

\usepackage{graphicx}
\usepackage{dcolumn}
\usepackage{bm}

\begin{document}

\title{Aging, rejuvenation and memory effects in re-entrant ferromagnets\\ }

\author{V. Dupuis, E. Vincent, M. Alba and  J. Hammann}

\affiliation{Service de Physique de l'Etat Condens\'e, DRECAM, DSM, CEA Saclay, 91191 Gif sur
Yvette Cedex, France }


\begin{abstract}
We have studied the slow dynamics of the ferromagnetic phases of the re-entrant
$CdCr_{2x}In_{2-2x}S_4$ system for $0.85<x \le 1$ by means of low frequency ac susceptibility and
magnetization measurements.  Experimental procedures widely used in the
investigation of the out-of-equilibrium dynamics of spin glasses (such as the $x=0.85$ compound)
have been applied to search for aging, rejuvenation and memory effects, and to test their
dependence on the disorder introduced by dilution of the magnetic ions. Whereas the rejuvenation
effect is found in all studied samples, the memory effect is clearly enhanced for increasing
dilutions. The results support a description 
of aging in both ferromagnetic and re-entrant spin-glass phases 
in terms of hierarchical reconformations of domain
walls pinned by the disorder. 
\end{abstract}
\pacs{75.40. Gb, 75.50.Lk}

\maketitle

\section{Introduction}
Disordered magnetic systems have been intensively studied in the last decades. Among them, the
systems with competing magnetic interactions are of particular interest. When the contributions of
ferromagnetic and antiferromagnetic interactions are comparable, conventional long range order is
not stable and a spin glass behavior is obtained. If the ferromagnetic interactions are
preponderant, the system displays a re-entrant behavior, with, for decreasing temperatures, a
transition from the paramagnetic phase to a ferromagnetic phase, followed by a transition to a spin
glass phase.

Experimentally, in some compounds the amount of ferromagnetic and antiferromagnetic interactions
can be continuously varied by changing the concentration of a magnetic
ion \cite{Maletta,Nogues,Nogues2}. 
One such example is the
$CdCr_{2x}In_{2-2x}S_4$ thiospinel {\cite{Nogues}}, a frustrated magnetic insulator with
ferromagnetic nearest neighbor and antiferromagnetic next nearest neighbor interactions. The
$x=0.85$ compound is a model spin glass in which aging has been extensively studied \cite{Sitges}.
The interest in aging phenomena has been revived by experiments on the effect of temperature
changes, which have revealed non trivial rejuvenation and memory effects \cite{MemChaos}, also
identified later on in many disordered systems \cite{KTN, Weissman, Ciliberto,Parker}. While these
features are well accounted for in hierarchical pictures where one assumes a hierarchical
organization of the metastable states of the spin glass as a function of temperature
\cite{Hierarchy}, their interpretation in the physical space \cite{Domains} is less clear and is
currently the subject of many efforts, both on the theoretical and experimental sides
\cite{Review,2dXY,Lemaitre,NumRecent,SGFerro}.

In this paper, we present a comparative study of spin-glass type phenomena, and in particular of
the rejuvenation and memory phenomena, in the ferromagnetic phases of the $CdCr_{2x}In_{2-2x}S_4$
system for several concentrations $x$ of the magnetic ions ($x>0.85$). Our motivation for studying
these ferromagnetic phases is the fact that since they are conceptually simpler than spin glasses
(spins are organized in domains), the rejuvenation and memory phenomena should be more easily
described in terms of geometrical spin arrangements. This argument already led us to study the
ferromagnetic phase of the weakly diluted $x=0.95$ compound \cite{SGFerro} in which we could
observe some rejuvenation and memory effects. The latter were discussed in terms of pinning and
reconformations of ferromagnetic domain walls in the dilution-induced disorder. The work that we
present here, involving the study of a more diluted sample ($x=0.90$), complements the previous
investigation and provides further insights on the role of disorder in these out-of-equilibrium
phenomena.

The paper is organized as follows. First, we present the phase diagram of the
$CdCr_{2x}In_{2-2x}S_4$ system, the general behavior of the magnetization and the low frequency ac
susceptibility of the studied samples with $x>0.85$. Then we analyse the results of isothermal ac
susceptibility relaxations at various temperatures both in the ferromagnetic phases and in the
re-entrant spin glass phases. This allows us, for example, to highlight the effect of disorder on
isothermal aging. We describe experiments in which we have investigated the effect of temperature
changes on aging. The comparison of the results for different concentrations $x$ shows that the
ability of a disordered ferromagnet to memorize aging at a given temperature is strongly dependent
on the disorder strength. Whereas for weak disorder ($x$ = 0.95), the memory of an isothermal aging
is easily erased by a short low temperature excursion (below the temperature at which the system
was aged) \cite{SGFerro}, for stronger disorder ($x$ = 0.90), this memory is preserved and can only
be erased by an excursion down to the re-entrant spin glass phase. Finally, we discuss the whole
set of results in terms of pinning and reconformations of the ferromagnetic domain walls in their
disordered environment and raise the question of the relevance of this wall dynamics for explaining
the rejuvenation and memory phenomena in spin glasses.

\section{Characterization of the system}
The investigated samples are chromium thiospinels of formula $CdCr_{2x}In_{2-2x}S_4$ for $x=1.00$,
$x=0.95$, $x=0.90$ and $x=0.85$ prepared by M. Nogu\`es at the CNRS Meudon-Bellevue laboratory.
They have been well characterized in the past
 by various techniques \cite{Nogues} including neutron diffraction \cite{Pouget} and display behaviors ranging from pure
ferromagnet, to re-entrant ferromagnet and spin glass for decreasing $x$. The samples are
polycrystalline powders and the measurements reported here were performed with a commercial
$Cryogenic^{Ltd}$ S600 SQUID magnetometer.

\subsection{Structure and phase diagram}
The phase diagram of the $CdCr_{2x}In_{2-2x}S_4$ is shown in Fig. 1. The pure compound
$CdCr_{2}S_{4}$ is a normal spinel, i.e. all the Cr$^{3+}$ ions occupy octahedral (B) sites of the
lattice whereas the tetrahedral (A) sites are occupied by Cd$^{2+}$  ions only. The Cr$^{3+}$ ions
are in a $3d^{3}$ electronic state, leading to a magnetic moment of spin only type with S=$3\over
2$. The main magnetic interactions between a Cr$^{3+}$ ion and its six nearest neighbors are
positive {\cite{Pouget}}  (${J_{1} / k_{B}}= 13.25 \pm 0.12 K$) and induce ferromagnetism below
$T_{C}=84.8~K$. The existence of super exchange antiferromagnetic interactions \cite{Dwight67}
between the twelve third nearest neighbors (${J_{3} / k_{B}}= -0.915 \pm 0.015 K$) induces
frustration in the magnetic lattice  that can be quantified  by the weighted interaction ratio $
R={ z_{3}\times J_{3}\over z_{1}\times J_{1}}= -0.138 $, to be compared to the critical ratio for
the stability of ferromagnetism $ R_{C} = -0.25 $ in the spinel structure \cite{Balt66}. The
dilution of the Cr$^{3+}$ ions by non magnetic In$^{3+}$ ions enhances this frustration as the
antiferromagnetic couplings via the In$^{3+}$ orbitals $J_{L}$ are amplified by a factor
$R_{In}=J_{L}/J_{3} \sim 9.0$ \cite{Veill86,Mery85} . This large enhancement explains
quantitatively the absence of a percolating ferromagnetic cluster for dilutions ($1-x$) greater
than about 15\% \cite{Nogues}. The dilute samples ($0.15 < x < 0.85$) therefore undergo a unique
phase transition towards a spin glass phase at low temperatures.

For higher concentrations in chromium ions ($0.85<x<1$), the sequence of phase transitions upon
cooling down is as followed: a para-ferromagnetic phase transition first occurs at a critical
temperature $T_{c}$, followed by the onset of spin glass type irreversibilities, at a lower
critical temperature $T_{g}$. This phenomenon is usually referred to as reentrance.

\begin{figure}[htbp]
\begin{center}
\centering {\par \resizebox*{!}{5.5cm}{\rotatebox{-90}{\includegraphics{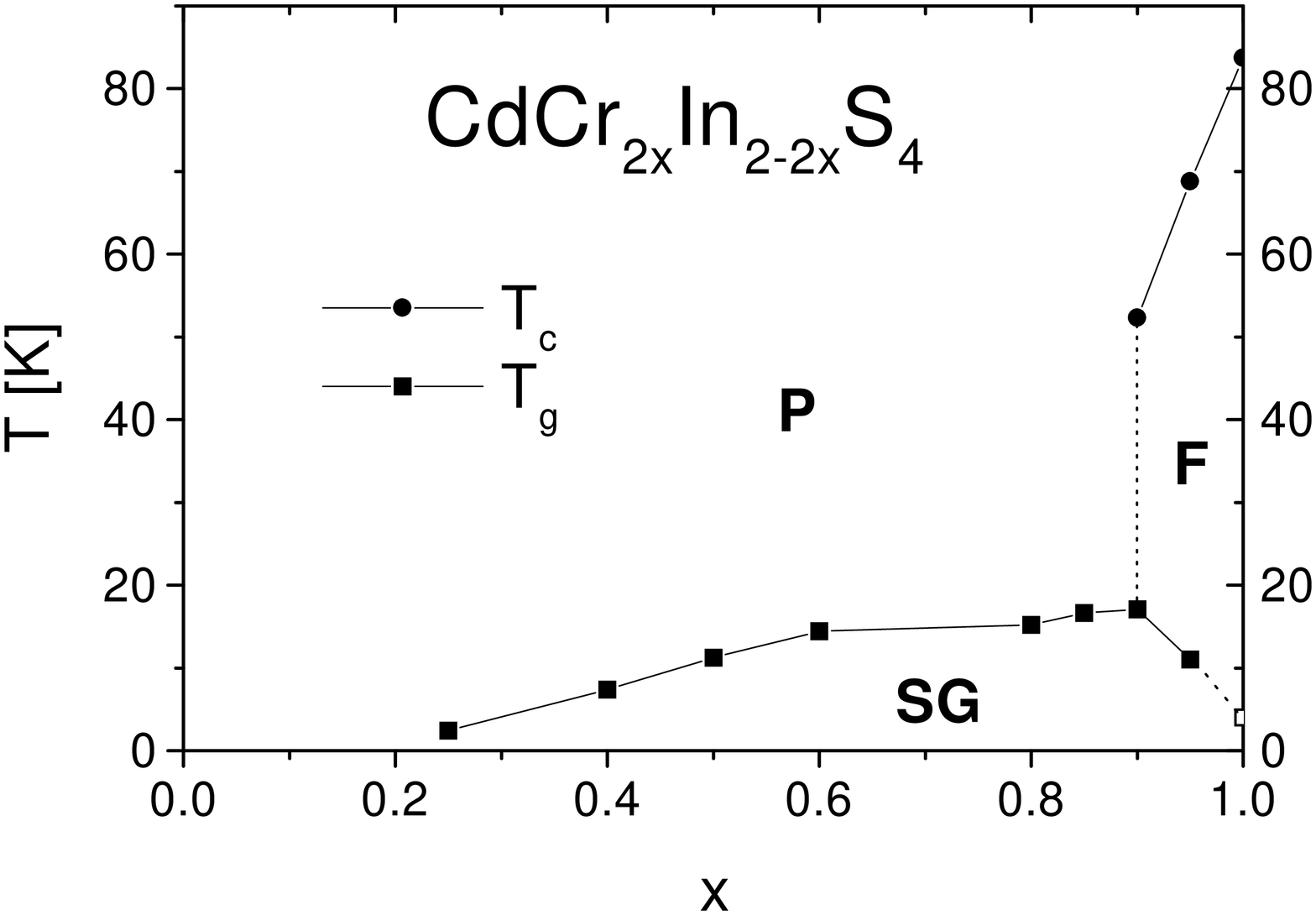}}}}
\end{center}
\caption{\label{Pdiag} Phase diagram of the $CdCr_{2x}In_{2-2x}S_4$. The points corresponding to
$x<0.85$ are extracted from \cite{Nogues}. For $x>0.85$, $T_c$ corresponds to the inflection point
of the increase of the out-of-phase susceptibility $\chi''(T)$ at low frequency when cooling from
high temperatures. Same convention for $T_g$. The open square for the $x=1.00$ sample corresponds
to the small $\chi''$ peak observed at low temperatures (see text). }
\end{figure}

\subsection{Field Cooled and Zero Field Cooled magnetizations}

Fig. 2 displays the Field Cooled (FC) and Zero Field Cooled (ZFC) magnetizations (normalized to the
value of the applied field) vs. temperature for the studied samples. 
The ZFC measurements have been performed after cooling the samples in zero
field (at $\sim 80\  mK/s$) to 3K, applying the field at 3K, and
recording the data upon slowly re-heating (at $\sim 3\  mK/s$). The FC
data have been taken in the same manner (but cooling {\it with the
field on}). The amplitude of the small probing field was typically of
$\sim 10\  Oe$ in order to remain in the linear response regime. We have
applied demagnetizing field factor corrections to the data to take into account the different
geometries of the samples \cite{demagnet}. The results are presented for a spherical shape factor.

While at high temperatures, all samples are paramagnetic, two types of behavior appear at lower
temperatures.

In the pure ($x=1.00$) and the weakly diluted ($x=0.95$) samples, a transition from the disordered
paramagnetic state to a long range ferromagnetic ordered state is signalled by a sharp increase of
both the ZFC and FC magnetizations above the Curie temperature $T_{c}$. This transition is
confirmed by neutron diffraction measurements which evidence spontaneous magnetization, magnetic
Bragg peaks and characteristic spin waves \cite{Pouget}. Below $T_c$, magnetic irreversibilities
are observed in the separation of the FC and ZFC curves: the magnetization becomes history
dependent. The slow increase of the FC magnetization as the temperature is decreased is accounted
for by a small extra polarization of the sample when cooled in a field and can be related to the
temperature dependence of the spontaneous magnetization. On the other hand, the plateau of the ZFC
magnetization, at a level equal to the inverse of the demagnetizing field factor, indicates that
the spins are well organized in ferromagnetic domains.

The behavior of the strongly diluted ($x=0.90$) sample is markedly different. The increase of the
FC and ZFC magnetizations upon cooling from the paramagnetic phase is less pronounced than in the
less diluted samples. Neutron diffraction measurements reveal only short range ferromagnetism with
no magnetic Bragg peaks nor a true divergence of the spin-spin correlations \cite{Pouget}. We can
still define for convenience a `transition' temperature $T_c$ at the inflection point of the
magnetization increase but it is important to note that this is not a true transition to a long
range ordered state. With this definition, we see that, just below $T_c$, no clear separation of
the FC and ZFC curves is found within experimental accuracy. Nevertheless, we observe a rounded
magnetization plateau at a level corresponding to the expected demagnetizing field factor which
suggests that the magnetic moments in the sample are still organized in ferromagnetic domains.

At lower temperatures, while nothing special happens in the pure sample, the FC magnetization of
the $x=0.95$ and $x=0.90$ samples decreases and tends to saturate. In parallel, we observe a sharp
drop of the ZFC magnetization. This change of behavior corresponds to the transition from the
ferromagnetic state to the strongly frozen re-entrant spin glass state and is to be compared with
the reference spin glass behavior of the more diluted $x=0.85$ sample also shown in Fig. 2.

\begin{figure}[htbp]
\begin{center}
\centering {\par \resizebox*{!}{5.5cm}{\rotatebox{-90}{\includegraphics{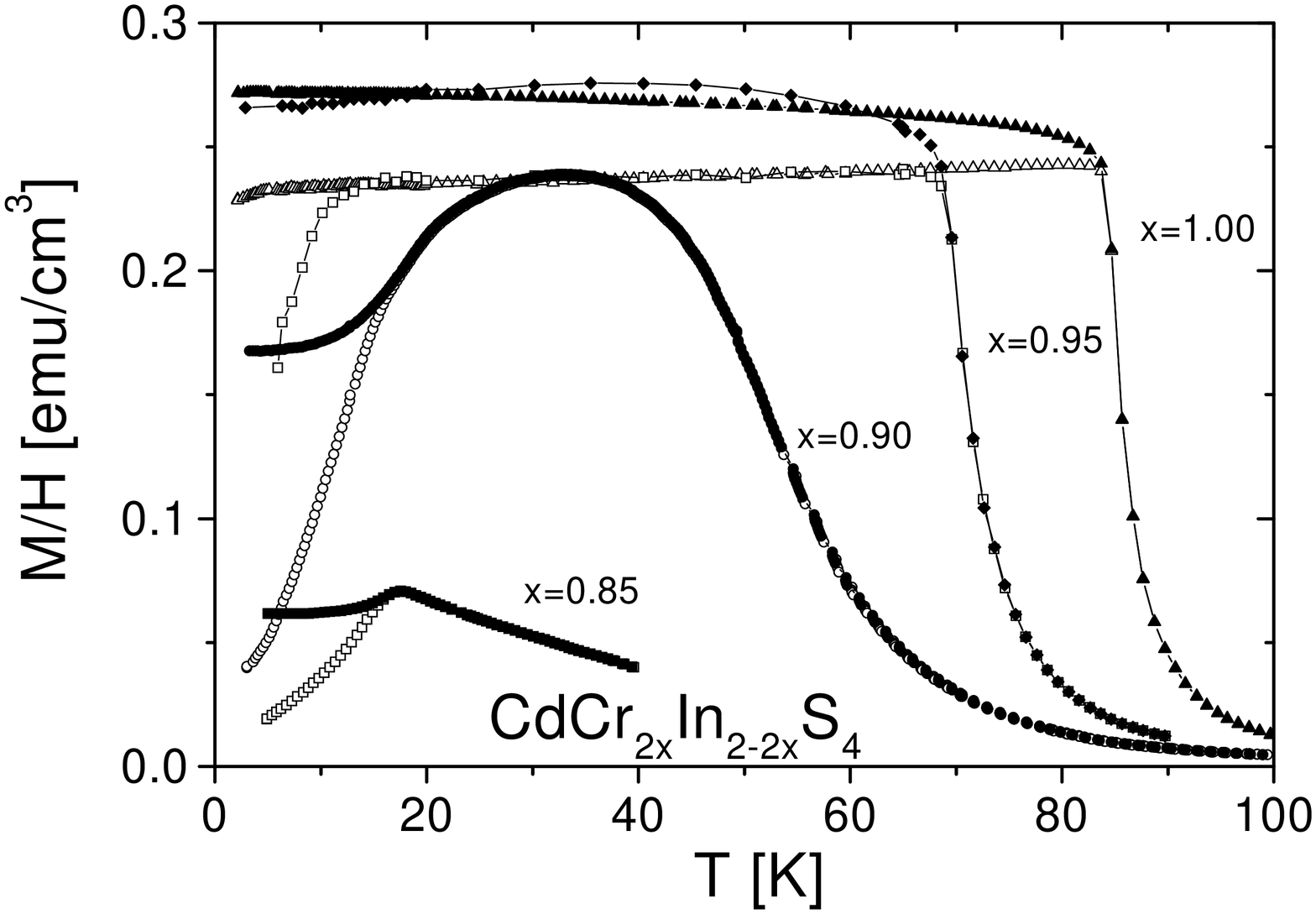}}}}
\end{center}
\caption{\label{ZFCFC} ZFC (open symbols) and FC (filled symbols) magnetizations divided by the
amplitude of the probing field vs. temperature of the $CdCr_{2x}In_{2-2x}S_4$ system, for various
concentrations $0.85 \le x \le 1$.}
\end{figure}

\begin{figure}[htbp]
\begin{center}
\centering {\par \resizebox*{!}{5.5cm}{\rotatebox{-90}{\includegraphics{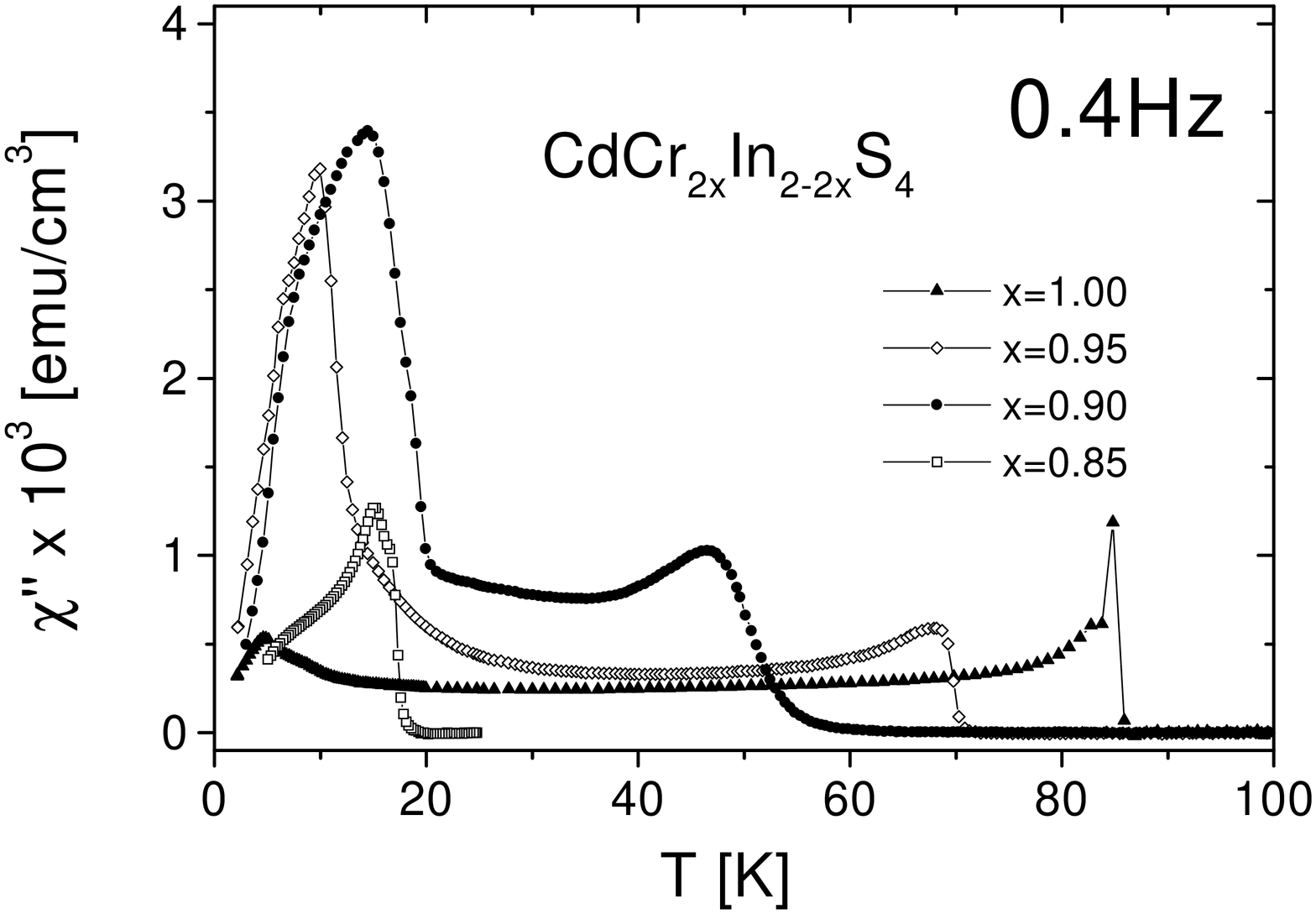}}}}
\end{center}
\caption{\label{chi2T} Temperature dependence of the out-of-phase susceptibility $\chi''$ of the
$CdCr_{2x}In_{2-2x}S_4$ system at $0.4 Hz$, for various concentrations $0.85 \le x \le 1$.}
\end{figure}

\subsection{Low frequency ac susceptibility}
To complete the previous analysis, we have measured the low frequency ac susceptibility
$\chi=\chi'-i\chi''$ of the samples as a function of temperature. Here
again, the data has been taken while re-heating at $\sim 1-2\ mK/s$,
after cooling  to 3K at $\sim 80\ mK/s$, using a
probing field of weak amplitude (typically 1.7 Oe peak value) in order to remain in the linear
response regime. We have used the paramagnetic phase response (assuming $\chi''=0$) to correct a
slight frequency dependent phase shift in the detection setup. In the following, we focus on the
out-of-phase component $\chi''$ which is directly related to the dissipation processes.

Fig. 3 shows $\chi''$ at $0.4Hz$ vs. temperature for all the studied samples. The onset of
ferromagnetism is signalled by a sharp increase in $\chi''$ when cooling from high temperatures.
This increase in the dissipation indicates that the characteristic relaxation times of the system
have reached a value comparable to the macroscopic measurement time (equal here to the inverse of
the ac field frequency). This is the standard scenario expected at a second order phase transition
where the average relaxation time diverges. We can define a transition temperature $T_c$ at the
inflection point of the $\chi''$-increase. We find that $T_c$ slightly decreases for decreasing
frequencies, in agreement with the expectation from dynamic scaling arguments. Below $T_{c}$,
$\chi''$ keeps a non zero value, roughly temperature independent in the whole ferromagnetic region
meaning that there are still dissipation processes at the time scale of the experiments in the
ferromagnetic phase of the studied samples, even in the pure one. The stronger the dilution, the
higher this plateau value of $\chi''$. The primary effect of an increase in the disorder strength
is therefore an increase of the level of low frequency dissipation.

At lower temperatures, sharp peaks with a critical frequency
dependence \cite{Tholence,Mydosh} are observed in the
out-of-phase susceptibility $\chi''$ of the diluted samples ($x=0.95$ and $x=0.90$) and signal the
transition from the ferromagnetic state to the re-entrant spin glass state. The corresponding
transition temperatures estimated from the inflection points at low frequency ($0.4Hz$) are about
the same as the ones at which strong irreversibilities are observed in FC-ZFC measurements.
Curiously, a small peak in $\chi''$ is also observed at low temperatures in the pure $x=1.00$
sample where no existence of a re-entrant spin glass phase has ever been reported. However, while
the frequency dependence of the peaks observed in the diluted samples is characteristic of critical
slowing down (in the frequency range explored [0.04Hz-8Hz]), we have found that the position of the
peak in the pure sample has an Arrhenius type frequency dependence, evocative of a dynamical
freezing of the spins. Even if the origin of this peak is at this moment not completely clear, it
does not seem to correspond to a phase transition as in the diluted samples.

\section{Slow dynamics and aging at constant temperature}
We have investigated the isothermal slow dynamics and aging of the $CdCr_{2x}In_{2-2x}S_4$ samples
in the following way. The samples were cooled from their paramagnetic phase down to various
successive aging temperatures $T_m$, in both the ferromagnetic and the re-entrant spin glass phases
(with a cooling rate of typically $1 mK/s$). At each fixed aging temperature, the time evolution of
the ac susceptibility was recorded for three different frequencies: $0.04Hz$, $0.4Hz$ and $4Hz$.

In all cases, we observe a slow time decay (aging) of the ac susceptibility $\chi(\omega , t)$
towards an asymptotic frequency dependent value $\chi_0(\omega)$ (stationary susceptibility). The
relaxation is more important in relative value on the out-of-phase component than on the in-phase
component and in the following we focus on $\chi''$.

The influence of temperature on aging is illustrated in table I, in which we report, for the
$x=0.95$ and $x=0.90$ samples, the amplitude of the $\chi''$ relaxations (in absolute and relative
values) at fixed frequency, observed at various temperatures and for a given waiting time
$t_w=15000s$. The trend is the same for both samples. The relaxations are always weaker in the
ferromagnetic phase than in the re-entrant spin glass phase. Moreover, in the ferromagnetic phase,
the relaxation is mainly important near $T_c$ and tends to disappear in the experimental background
noise at lower temperatures.

Concerning the frequency dependence of the relaxation, we observe that the lower the frequency
$\omega$, the greater the relaxation amplitude observed in the experimental time window. This
result is in agreement with a previous investigation of the $x=0.95$ sample {\cite{SGFerro}}, where
a qualitative $\omega t$ scaling (characteristic of spin glasses \cite{Sitges}) of the non
stationary part of the susceptibility was found both in the re-entrant spin glass and in the
ferromagnetic phases. Here, we have studied this scaling behavior of the isothermal ac
susceptibility relaxation in a more quantitative way. We have recorded the relaxation of the ac
susceptibility of the $x=0.90$ sample following a quench ($\sim 80\
mK/s$) to a temperature $T_m=42K$ in the
ferromagnetic phase for $8$ frequencies $\omega$ of the ac field in the range [$0.04Hz$-$4Hz$].
Fig. 4 shows the resulting relaxation curves for the out-of-phase component $\chi''$ as a function
of the reduced variable $\omega t$. As we can see in the figure, the scaling works pretty well over
nearly five decades. All the relaxation curves, after subtraction of a constant which corresponds
to the stationary part $\chi''_{0}(\omega)$, lie on a unique master curve which is a function of
$\omega t$ and can be nicely fitted by a power law with an exponent of order $-0.2$ (same value as
commonly observed in spin glasses {\cite{Sitges}}).

\begin{table}
\caption{Amplitude of $\chi''$ relaxation at 0.4 Hz after slow cooling
($\sim 1\ mK/s$), 
in absolute and relative values, 
for the $x=0.95$ and the $x=0.90$ $CdCr_{2x}In_{2-2x}S_4$ samples after a waiting time $t_w = 15000 s$.}
\begin{ruledtabular}
\begin{tabular}{lcc}
T [K] & $\Delta \chi''$x$ 10^5$ [emu/cm$^3$] & $\Delta \chi''/\chi''$x$100$\\
\hline
x=0.95\footnote{$T_c=70K$ and $T_g=10K$.} & & \\
\hline
67 (0.95$T_c$) & 2.4 & 4.7\\
40 (0.57$T_c$) & 1.0 & 3.1\\
8 (0.80$T_g$) & 26.2 & 11.6\\
5 (0.50$T_g$) & 8.6 & 5.3\\
\hline
x=0.90\footnote{$T_c=50K$ and $T_g=18K$.} & & \\
\hline
42 (0.84$T_c$) & 3.4 & 3.8\\
30 (0.60$T_c$) & 1.1 & 1.6\\
17 (0.95$T_g$) & 16.0 & 6.7\\
10 (0.56$T_g$) & 29.0 & 12.9\\
\end{tabular}
\end{ruledtabular}
\end{table}

\begin{figure}[htbp]
\begin{center}
\centering {\par \resizebox*{!}{5.5cm}{\rotatebox{-90}{\includegraphics{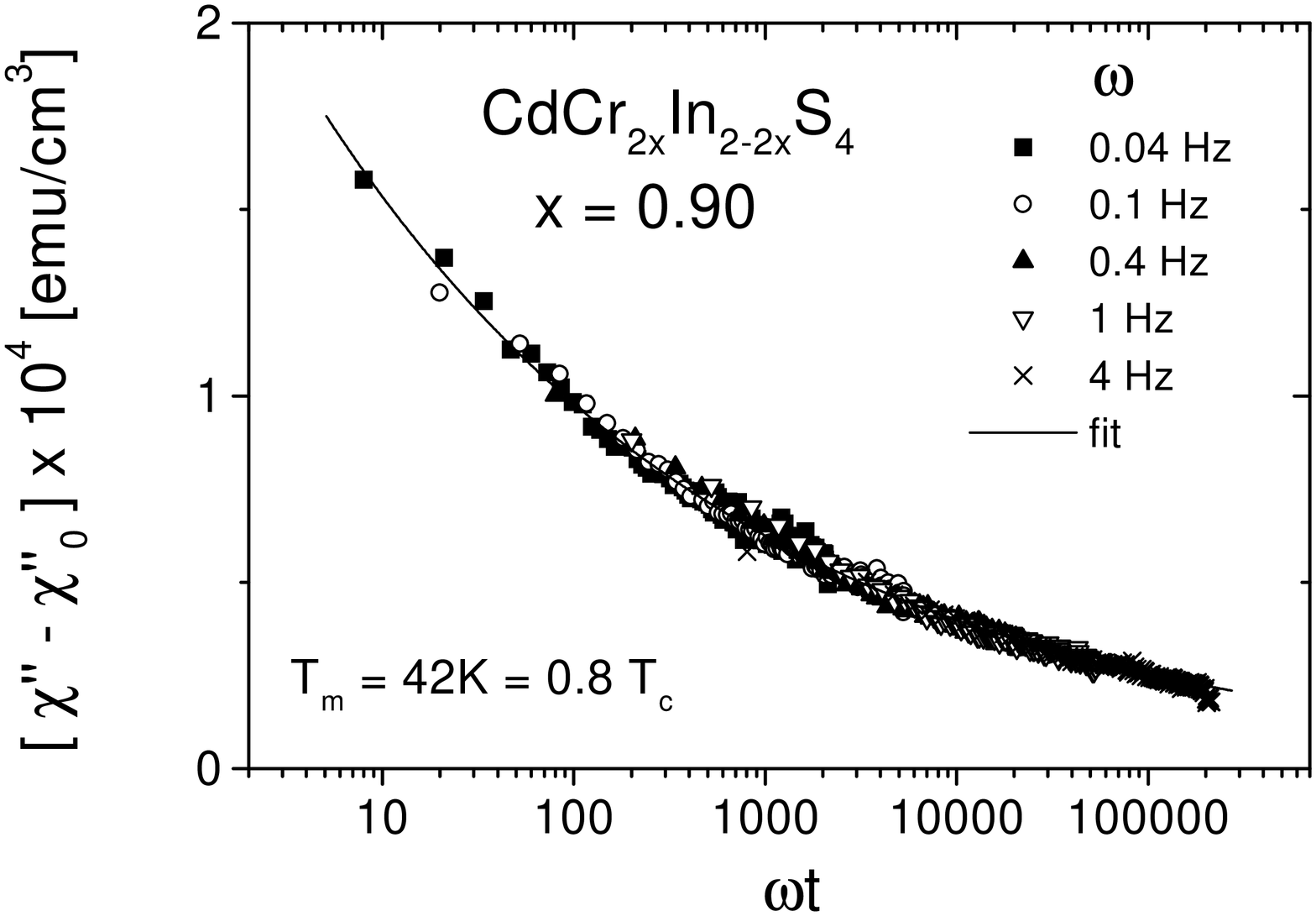}}}}
\end{center}
\caption{\label{scaling} Scaling plot of the $\chi''$ relaxation
curves of the $x=0.90$ sample after a quench ($\sim 80\ mK/s$) to 
$T_m=42K$ as a function of $\omega t$ and for frequencies of the ac field ranging from 0.04 Hz to
4Hz. The asymptotic stationary part $\chi''_0$ has been subtracted. All the curves lie on an unique
master curve which can be nicely fitted by a power law with a small exponent -0.2. }
\end{figure}

\section{Rejuvenation and memory effects}

In conventional spin glasses, such as the $x=0.85$ compound, it is known that decreasing the
temperature from T to T-$\Delta$T in the spin glass phase, after having isothermally aged the
sample at T, strongly restarts the dissipation processes. This is the rejuvenation effect
\cite{MemChaos}, the system seems to forget its previous equilibration stage at T. For a
sufficiently large $\Delta$T, the system rejuvenates completely and the relaxation of the ac
susceptibility at T-$\Delta$T is identical to that obtained after a direct quench from the
paramagnetic phase. The other striking feature happens when the system is re-heated to the initial
temperature T after a given low temperature history. No restart of aging is found and the
relaxation is the exact continuation of the one that occurred before going to low temperatures.
This is the memory effect \cite{MemChaos}, aging at T-$\Delta$T had no influence on aging at T.
These effects have been interpreted as an evidence for a hierarchical organization (tree-like for
example) as a function of temperature of the metastable states of the spin glass
{\cite{Hierarchy}}.

In order to reach a better understanding of these non trivial rejuvenation and memory effects in
terms of geometrical spin arrangements, we have previously investigated the pure and weakly
disordered ferromagnetic phase of the $x=1.00$ and $x=0.95$ compounds \cite{SGFerro}. We found
clear rejuvenation effects coexisting with a very weak memory of aging, which was easily erased by
an excursion at lower temperature. Here, we report similar investigations of the
rejuvenation and memory phenomena in the short range ferromagnetic phase of the $x=0.90$ sample.
Our results shed light on the role of disorder in the rejuvenation and memory phenomenology.

\begin{figure}[htbp]
\begin{center}
\centering {\par \resizebox*{!}{5.5cm}{\rotatebox{-90}{\includegraphics{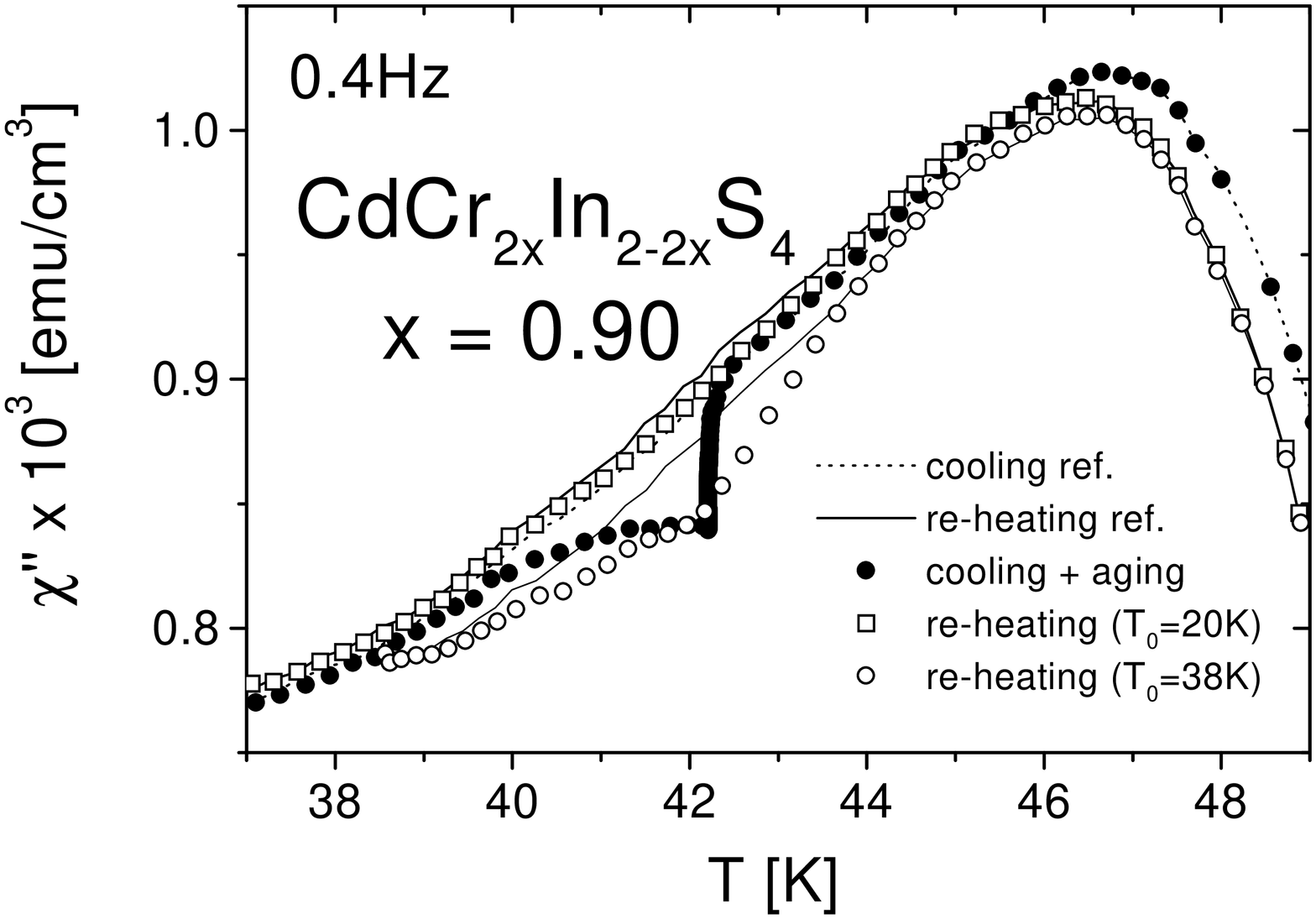}}}}
\end{center}
\caption{\label{rejmemtemp} Aging, rejuvenation and memory effects on $\chi''$ vs. temperature in a
single stop experiment at $T_m=42K$ (see details about the procedure in the text). Aging is
evidenced by the downward relaxation of $\chi''$ while the rejuvenation effect corresponds to the
increase of $\chi''$ back to the reference when cooling is resumed down to $T_0$. Upon re-heating,
a clear memory dip centered around $T_m$ is seen for $T_0=38K$. In contrast, for $T_0=20K$, no
memory is found. }
\end{figure}

In a first experiment, we have studied the rejuvenation and memory phenomena in the temperature
dependence of the ac susceptibility. Fig. 5 shows the out of phase susceptibility $\chi''$ of the
$x=0.90$ sample at $0.4Hz$ vs. temperature. All along the experiment a sweeping rate $|dT/dt|$ of
$\sim 3\ mK/s$ was used. The sample was cooled down to a measurement temperature $T_m=42K$ where it was
isothermally aged during $8h$ (full symbols). After this aging stage, the cooling was resumed down
to a final temperature $T_{0}$ (two cases, $T_{0}=38K$ and $T_{0}=20K$, were studied) and finally,
the sample was slowly re-heated from $T_{0}$ to the paramagnetic phase (open symbols). For
comparison, a reference curve, recorded with the same protocol but without the aging stage, is also
plotted on this graph (dotted and solid lines for cooling and re-heating).

Aging is clearly visible in the downward relaxation with time of $\chi''$ at $T_m$. When cooling is
resumed after the aging stage at $T_m$, a rejuvenation effect is evidenced by the increase of
$\chi''$ back to the reference cooling curve. Considering now the re-heating process, we see that
in the experiment with a short excursion down to $T_{0}=38K$, a memory of the aging at $T_{m}$ is
found as $\chi''$ departs from the heating reference curve on approaching $T_m$ and displays a
characteristic dip around $T_{m}$ before merging back with the reference curve. In contrast, in the
experiment with a longer excursion down to $T_{0}=20K$, the re-heating curve remains superimposed
onto the re-heating reference curve and no memory dip is found. Let us mention that in another
experiment made with the same protocol but with $T_{0}=30K$, we observed a memory effect identical
to the one observed with $T_{0}=38K$.

\begin{figure}[htbp]
\begin{center}
\centering {\par \resizebox*{!}{5.5cm}{\rotatebox{-90}{\includegraphics{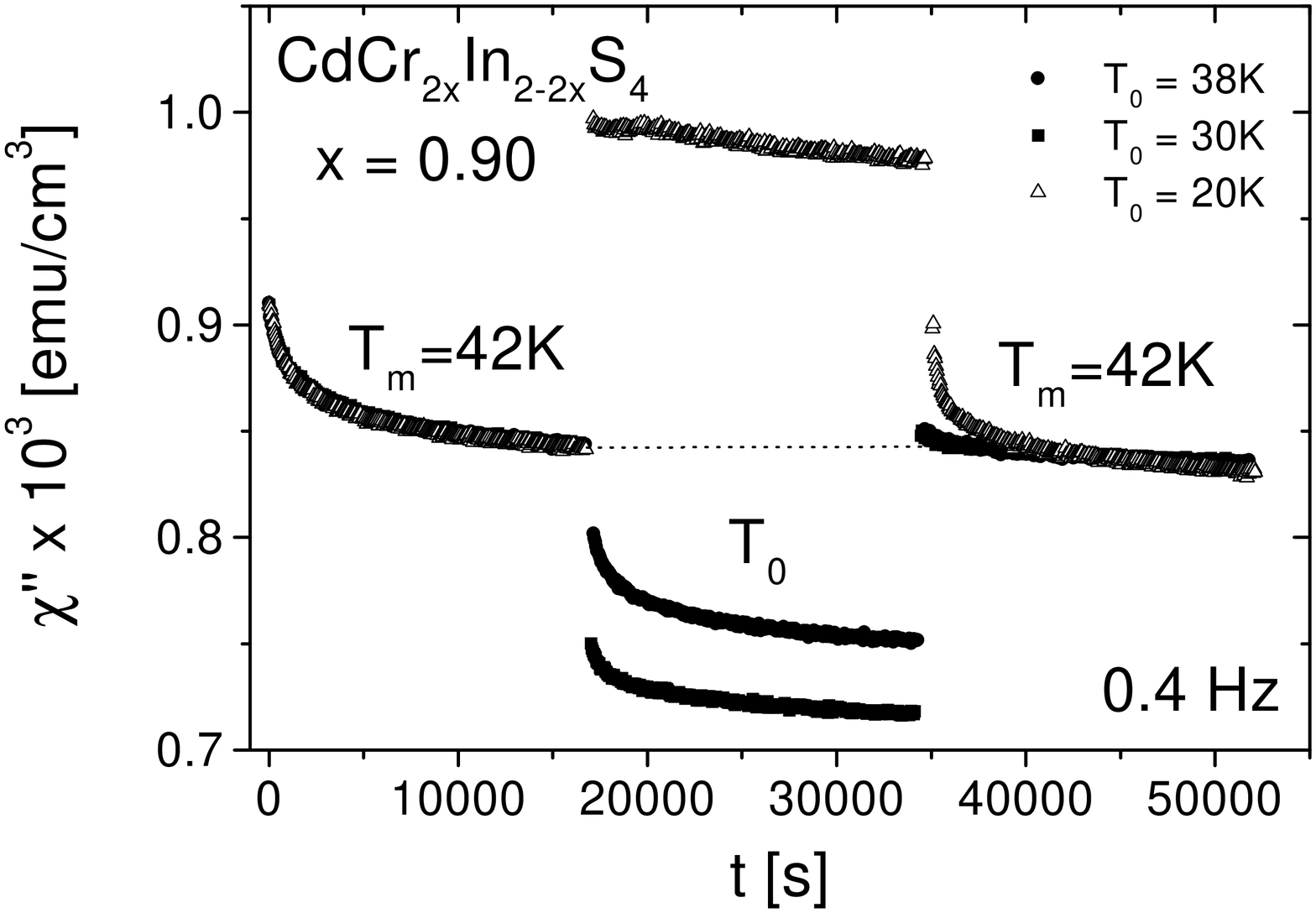}}}}
\end{center}
\caption{\label{rejmemtime} Aging, rejuvenation and memory effect on $\chi''$ vs. time in
temperature cycling experiments. After the quench from high temperature down to $T_m=42K$, $\chi''$
is relaxing downwards with time. Cooling down after $17000s$ to $T_0$, induces a rejuvenation effect
evidenced here by the restart of relaxation processes for $T_0=38K$ and $T_0=30K$, not for
$T_0=20K$. When re-heating to $T_m$, the relaxation of $\chi''$ is the continuation (memory) of the
one that occured before the cycle for $T_0=38K$ and $T_0=30K$. In contrast, for $T_0=20K$ a strong
restart of aging processes, analogous to a quench from high temperature is seen meaning that no
memory is found is this case. }
\end{figure}

In a second experiment, we have studied the rejuvenation and memory phenomena as a function of
time. Fig. 6 shows the out of phase susceptibility $\chi''$ at $0.4Hz$ vs. observation time
measured with a negative temperature cycling protocol : the sample was quenched from high
temperatures down to the measurement temperature $T_m=42K$, aged there during $t_1=17000s$, cooled
then to $T_{0}$ , aged at this temperature another $t_2=17000s$ then heated back to $T_m$ and
finally aged there again for $t_3=17000s$. Three values were used,  $T_{0}= 38K$, $30K$ and $20K$.

Just after the quench, $\chi''$ relaxes downwards with time (it is the same kind of relaxation as in
the exp. of Fig. 5). Cooling from $T_m$ to $T_{0}$ induces a restart of aging for $T_{0}=38K$ and
$T_{0}=30K$: it is the rejuvenation effect observed here as a function of time. Heating back to
$T_m$ reveals the memory of the previous aging stage at $T_m$: except for a short transient regime,
the relaxation at $T_m$ is the exact continuation of the one that
occurred before the cycle.
However, in the experiment going down to $T_{0}=20K$ (that is, in the border region
between the ferromagnetic and spin-glass phases), the 
relaxation at  $T_{0}=20K$ does not show any upturn signing up a
rejuvenation effect.  We do not have a reference curve of a relaxation at
$20K$ after a direct quench for comparison; further studies in this
temperature region would be of interest. A  strong
restart of aging is observed when coming back to $T_{m}$ from
$T_{0}=20K$ after the negative cycle (in agreement
with the absence of memory in the experiments of Fig. 5). We have checked that this restart of
aging is not a long transient which would yield to a memory at a longer observation time as the
data points cannot be superposed onto a reference isothermal relaxation curve.

In summary, the experiments reported above show that the rejuvenation effect is a characteristic
feature of disordered ferromagnetic phases, as was already observed in {\cite{SGFerro,
FerroUppsChaos}}. It is not influenced by the amount of disorder. In contrast, the ability of a
ferromagnet to store and retrieve a memory of a previous aging strongly depends on the disorder
strength. Whereas the memory is weak and rapidly erased by a low temperature excursion for the
weakly disordered $x=0.95$ sample, it is much more robust in this more disordered $x=0.90$ sample.
For the latter, the memory of aging is not much affected by a low temperature excursion below $T_m$
as long as the lowest temperature reached during the experiment remains sufficiently above the
re-entrant spin glass transition temperature.

While performing the previous experiments on the $x=0.90$ sample, we observed a surprising behavior
of the ac susceptibility which may be related to the mechanism responsible for the erasure of the
memory. When recording the cooling/re-heating reference curves of Fig. 5 for $T_{0}=20K$, we found
that in the region [20K,35K], the cooling curve was always below the re-heating curve. This means
that the state which progressively develops in the short range ferromagnetic phase while cooling
from the paramagnetic phase is less dissipative than the one obtained when re-heating from the low
temperature re-entrant spin glass phase. This rather unusual phenomenon is displayed in Fig. 7. We
have investigated the aging properties of these two states which we call for convenience the cooled
state and the re-heated state. Fig.7 shows $\chi''$ vs. temperature at $0.4Hz$ measured along the
following procedure: slow cooling from high temperatures to $3K$ with an aging stop at $30K$ during
$4h$ (full circles), slow cooling to $3K$ and re-heating with an aging stop at $30K$ for $6h$ (open
circles) during the re-heating.

In the figure, the hysteresis between cooling and re-heating is clearly evidenced. If we consider
now the relaxations, we find a larger relaxation amplitude in the re-heated state than in the
cooled state. This extra dissipation gained in the vicinity of the re-entrant spin glass transition
indicate some incompatibility between the short range ferromagnetic correlations that develop while
cooling the sample from the paramagnetic phase and those that develop in the vicinity of the
re-entrant spin glass phase. The region where we observe a large hysteresis between cooling and
re-heating may be a region where the two types of correlations coexist in proportions that depend on
the thermal history. In the cooled state, this region corresponds mainly to a ferromagnetic pattern
whereas in the re-heated state, it corresponds mainly to a spin glass pattern. We believe that the
development of these spin glass correlations at low temperature is responsible for the erasure of
the memory of any previous aging in the short range ferromagnetic state.

\begin{figure}[htbp]
\begin{center}
\centering {\par \resizebox*{!}{5.5cm}{\rotatebox{-90}{\includegraphics{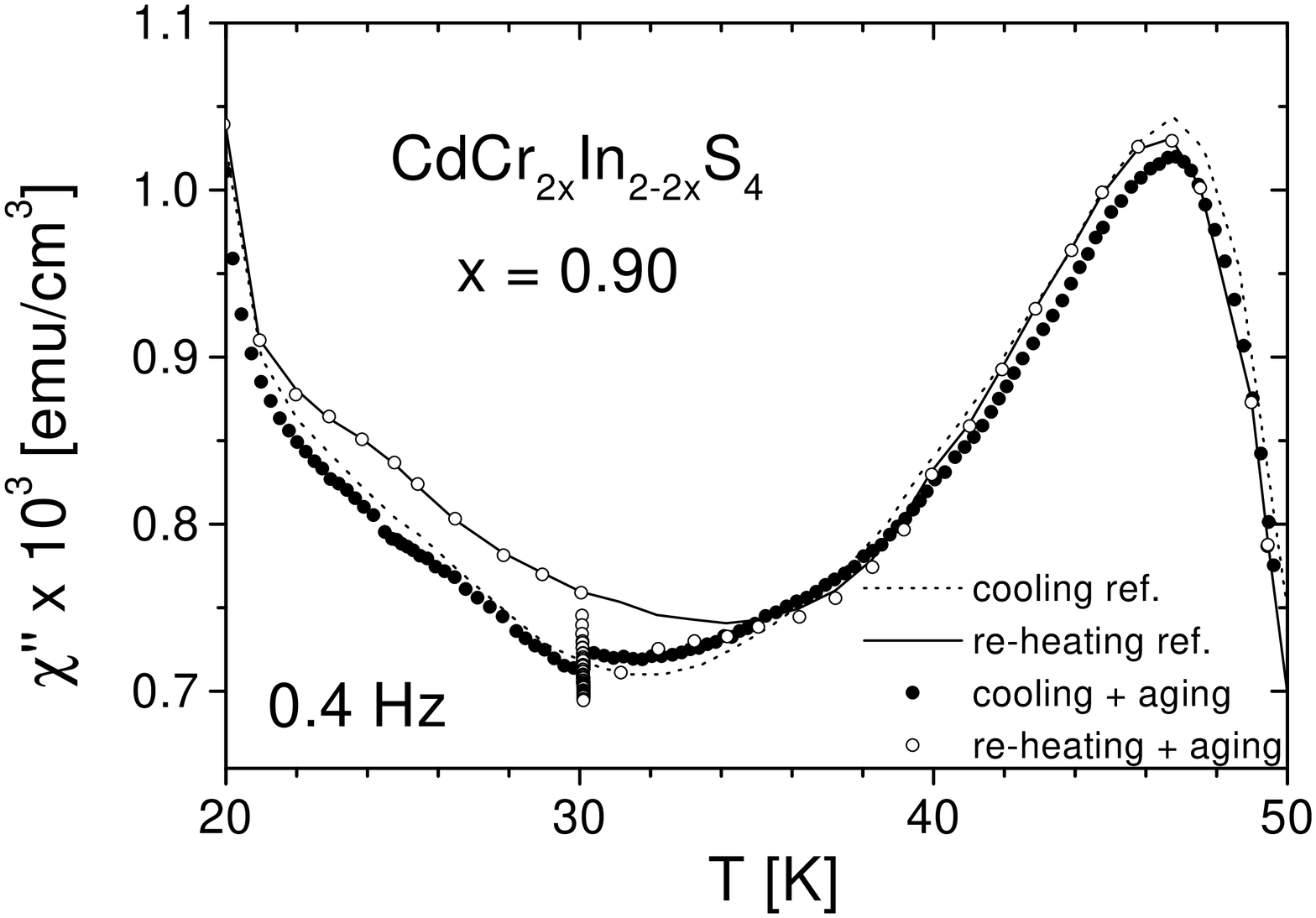}}}}
\end{center}
\caption{\label{anom} Relaxation of $\chi''$ during a waiting time $t_w$ at $T_m=30K$ for the
$x=0.90$ sample. Filled circles: the relaxation is made during the cooling, $t_w=4h$. Open circles:
the relaxation is made during the re-heating after continuous cooling down to $T_0=20K$, $t_w=6h$. The dotted and solid
lines correspond to cooling and re-heating reference curves (the
temperature sweeping rates are of order $1-4\ mK/s$). The (downwards) relaxation amplitude is larger in
the re-heated state than in the cooled state. The cooling and re-heating reference curves display
the same trend as the re-heating curve is clearly above the cooling one in the region [20K-35K].}
\end{figure}

\section{Discussion}
The series of experiments presented in this paper illustrates how the static and dynamic properties
of a ferromagnet, the $CdCr_2S_4$ are influenced by the introduction of site disorder. In a
non-disordered ferromagnet, the susceptibility is mainly related to the domain pattern in the
sample via the demagnetizing field, and the out-of-phase ac susceptibility gives more specific
informations on the stiffness and the density of the domain walls. In this case, dynamic properties
simply reflect the growth of the domains with time, which mainly proceeds by wall displacements.
This phenomenon which is only related to the surface tension of the walls is a fast process.

The introduction of pinning disorder, such as site dilution or structural defects, changes
drastically this picture \cite{Balents}. The frustration which arises from the competition between
the elastic energy associated with the wall deformation (which tends to favor a smooth interface)
and the pinning energy associated with the disorder (which favors a rough interface) results in a
drastic slowing down of dynamical processes. A domain wall has now many configurations between
which it can make thermally activated hops. The wall displacements hindered by the frustration then
proceed by reconformations of some parts of the wall which roughly correspond to jumps from one
pinning site to another. In the presence of a driving magnetic field which forces the domain growth
(a field ramp for exemple), this phenomenon is commonly known as the Barkhausen noise. In contrast,
without any driving field, the overall domain growth can be extremely slow and one can observe the
slow reconformations of the domain walls in a static pinning disorder without any notable growth of
the domains.

Actually, the dynamics of pinned elastic interfaces has many similarities with that of
spin-glasses. Theoretical arguments \cite{Balents} suggest that their energy landscape is
hierarchical with small length scale $l$ reconformations corresponding to hoping of small energy
barriers $E(l)\sim\Upsilon l^\theta$. In spin glasses, the rejuvenation effect upon cooling and the
memory when re-heating are usually ascribed to a hierarchical organization of the metastable states
\cite{Hierarchy}. When the temperature is lowered, the system remains in a deep well (memory
effect), while new subwells appear, inducing new aging processes (rejuvenation effect). A similar
hierarchical scheme is expected for the dynamics of a pinned domain wall. As time elapses at fixed
temperature, the wall equilibrates on longer and longer length scales (aging), while shorter length
scale processes are fluctuating in equilibrium. When the temperature is lowered, the conformation
of the wall on large length scale freezes (allowing for the memory upon re-heating) and the
equilibration now restarts on shorter length scales, which are no longer in equilibrium because
their Boltzmann weight has changed (rejuvenation). In parallel with these hierarchical
reconformation processes of the walls, the average domain size may grow with time. In that case,
the disorder seen by the domain wall is changing with time and obviously, this net motion tends to
erase the memory of the reconformations.

The above scenario accounts well for the observed behavior in the pure $x=1.00$ and weakly diluted
$x=0.95$ samples \cite{SGFerro}, and also that of some disordered dielectric crystals
{\cite{Levelut}}. In both $x=1.00$ and $x=0.95$ samples, aging and rejuvenation are clearly found
in parallel with a weak memory of aging which is erased by domain growth during a long excursion to
lower temperatures. The case of the more disordered $x=0.90$ sample, described here in details, is
interesting since it shows that the memory effect is stabilized by the disorder and supports the
scenario described above. The stronger the pinning of the domain walls, the slower their overall
displacement, the better the memory of the domain wall reconformations. This memory can
nevertheless be erased by a low temperature excursion close to the reentrant spin glass transition,
which indicates that the spin glass order establishing at low temperature is incompatible with the
short range ferromagnetic order developing above the re-entrant spin glass transition temperature.

It is tempting to extend this scenario of hierarchical wall reconformations in a pinning disorder
to the case of spin glasses, because it provides us with a simple explanation for the rejuvenation
and memory phenomena in the real space of spins without invoking temperature chaos arguments
\cite{Domains,Lemaitre} (see \cite{2dXY,Berthier}). In a spin glass, the nature of these walls is not as clear as
in a ferromagnet, and remains an open question. They may be related to the non trivial sponge-like
excitations recently discussed from the results of 3d spin-glass simulations \cite{Martin}. It
would be interesting to test the effect of temperature changes on such excitations.

Finally, let us mention that our results are consistent with other investigations on the re-entrant
ferromagnet $(Fe_{0.20}Ni_{0.80})_{75}P_{16}Al_{3}$ and also on the disordered manganite
$Y_{0.7}Ca_{0.3}MnO_3$ {\cite{FerroUppsChaos}}. The former sample ressembles our $x=0.90$ compound
regarding the temperature dependence of the susceptibility (rounded plateau in the ferromagnetic
region); the absence of memory effect is probably due to the proximity of the re-entrant spin glass
phase transition.

\section{Conclusion}
In this article, we have studied the slow dynamics of the ferromagnetic phases of the
$CdCr_{2x}In_{2-2x}S_4$ system for $x=1.00$, $x=0.95$ and $x=0.90$ by means of ac susceptibility
and magnetization measurements. We found that increasing the disorder strength induces a slowing
down of the ac response of the ferromagnetic phases resulting in an increase of the low frequency
dissipation. Using relaxation procedures developed for the study of the out of equilibrium dynamics
of spin glasses, we have found an aging behavior of the low frequency ac susceptibility both in the
ferromagnetic phases and in the re-entrant spin glass phases, with the same qualitative features as
in conventional spin glasses .

We have searched for rejuvenation and memory phenomena in the ferromagnetic phases of the studied
samples. Whereas the weakly disordered $x=0.95$ compound displays only a weak memory of aging
(rapidly erased by an excursion at lower temperatures), the more disordered $x=0.90$ compound shows
a more robust memory of aging which can only be erased by the growth of spin glass correlations in
the vicinity of the re-entrant spin glass transition. Our results support a scenario in which,
aging, rejuvenation and memory are described in terms of hierarchical reconformations of elastic
walls in a random pinning disorder. In this picture, the memory is contained in the large length
scale conformation of the walls and is erased by the growth of the domains. Our results show that
an increase in the disorder strength tends to reduce the domain growth and allows to observe clear
rejuvenation and memory effects. The extension of this wall reconformation scenario to the case of
spin glasses raises interesting questions such as the nature of the domains and the walls.

\vskip 1.0cm

\noindent {\it Acknowledgements}

\noindent We are grateful to X. Colson for his active participation to this work. We also wish to
thank J.-P. Bouchaud, F. Bert, D.H\'erisson, M. Ocio for stimulating discussions
during this work, L. Berthier and M. Nogu\`es for important remarks on the manuscript,
and L. Le Pape
for his technical assistance.


\vskip-12pt
\begin{thebibliography}{99}

\bibitem{Maletta} H. Maletta, {\it J. Appl. Phys.} {\bf 53}, 2185 (1982).

\bibitem{Nogues} M. Alba, J. Hammann and M. Nogu\`es, {\it J.
Phys. C}, {\bf 15}, 5441 (1982); J. L. Dormann, A. Saifi, V. Cagan and M. Nogu\`es, {\it Phys.
Stat. Sol. B} {\bf 131}, 573 (1985); M. Alba, J. Hammann, M. Ocio and Ph. Refregier, {\it J. Appl.
Phys.} {\bf 61}, 3683 (1987).

\bibitem{Nogues2} M. Nogues and J.L. Dormann, {\it J.M.M.M.} {\bf
54-57}, 87 (1986).

\bibitem{Sitges} E. Vincent, J. Hammann, M. Ocio, J.-P. Bouchaud,
L.F. Cugliandolo, in {\it Complex Behaviour of Glassy Systems}, Springer Verlag Lecture Notes in
Physics Vol.492, M. Rubi Editor, 1997, pp.184-219 (preprint cond-mat/9607224), and references
therein.

\bibitem{MemChaos} K. Jonason, E. Vincent, J. Hammann, J.-P.  Bouchaud
and P. Nordblad, {\it Phys. Rev. Lett.} {\bf 31}, 3243 (1998); K. Jonason, P. Nordblad, E. Vincent,
J. Hammann and J.-P. Bouchaud, {\it Europ. Phys. Jour. B} {\bf 13}, 99 (2000).

\bibitem{KTN} P. Doussineau, T. de Lacerda-Ar\^oso and A. Levelut, {\it Europhys. Lett.} {\bf 46}, 401 (1999).

\bibitem{Weissman} E. V. Colla, L. K. Chao, M. B. Weissman and D.D. Viehland, {\it Phys. Rev. Lett.} {\bf 85}, 3033 (2000).

\bibitem{Ciliberto} L. Bellon, S. Ciliberto and C. Laroche, {\it Europhys. Lett.} {\bf 51}, 551 (2000).

\bibitem{Parker} A. Parker and V. Normand, unpublished.

\bibitem{Hierarchy} Ph. Refregier, E. Vincent, J. Hammann and M. Ocio,
{\it J. Phys. France} {\bf 48}, 1533 (1987); E. Vincent, J.-P. Bouchaud, J. Hammann and F. Lefloch,
{\it Phil. Mag. B} {\bf 71}, 489 (1995).

\bibitem{Domains} D. S. Fisher and D. A. Huse, {\it Phys. Rev. B} {\bf
38}, 373 and 386 (1988); G. J. M Koper and H. J. Hilhorst, {\it J. Phys.  France} {\bf 49}, 429
(1988).

\bibitem{Review} For a review of different theoretical models, see: J.-P. Bouchaud, L. F.
Cugliandolo, J. Kurchan, M. M\'ezard, in `Spin-glasses and Random Fields', pp.161-224, A. P. Young
edt. (World Scientific, 1998).

\bibitem{2dXY} L. Berthier and P.C.W. Holdsworth, {\it
Europhys. Lett.} {\bf 58}, 35 (2002).

\bibitem{Lemaitre} H. Yoshino, A. Lema\^itre and J.-P. Bouchaud, {\it
Eur. Phys. J. B} {\bf 20}, 367 (2001).

\bibitem{NumRecent} H. Rieger, {\it Ann. Rev. of Comp. Phys. II},
ed. D. Stauffer (World Scientific, Singapore, 1995); M. Picco, F. Ricci-Tersenghi and F. Ritort,
{\it Phys. Rev. B}
 {\bf 63}, 17412 (2001), and {\it Eur. Phys. J. B} {\bf 21}, 211 (2001); T. Komori, H. Yoshino and H. Takayama, {\it J. Phys. Soc. Jpn.} {\bf
68}, 3387 (1999), {\it J. Phys. Soc. Jpn.} {\bf 69}, 1192 (2000), {\it J. Phys. Soc. Jpn.} {\bf 69}
Suppl. A, 228 (2000); E. Marinari, G. Parisi, F. Ritort and J. J. Ruiz-Lorenzo, {\it Phys. Rev.
Lett.} {\bf 76}, 843 (1996); A. Billoire and E. Marinari, {\it J. Phys. A} {\bf 33}, L265 (2000).

\bibitem{SGFerro} E. Vincent, V. Dupuis, M. Alba, J. Hammann and
J.-P. Bouchaud, {\it Europhys. Lett.} {\bf 50} (5), 674 (2000).

\bibitem{Pouget} S. Pouget, M. Alba, M. Fanjat and M. Nogu\`es,
{\it Physica B}, {\bf 180-181}, 244 (1992); S. Pouget and M. Alba, {\it J. Phys. Condens. Matter},
{\bf 7}, 4739 (1995).

\bibitem{Dwight67}
K. Dwight and N. Menyuk, Phys. Rev. {\bf 163},  435  (1967).

\bibitem{Balt66}
P.~K. Baltzer, P.~J. Wotjowicz, M. Robbins, and E. lopatin, Phys. Rev. {\bf
  151},  367  (1966); M. Nauciel-Bloch and R. Plumier, {\it Sol. Stat. Comm.} {\bf 9}, 223 (1971).

\bibitem{Veill86}
P. Veillet and K. Le-Dang, Phys. Rev. {\bf B} {\bf 33},  7855 (1986).

\bibitem{Mery85}
M. Mery, P. Veillet, and K. Le-Dang, Phys. Rev. {\bf B} {\bf 31}, 2656
  (1985).

\bibitem{demagnet} We have found that the measured susceptibilities in
the ZFC plateau region of the three samples are compatible with the
demagnetizing field factors corresponding to their overall macroscopic
shape. Therefore, the correction that we have applied to the
susceptibilities brings all ZFC plateau values to the same level
(corresponding to the demagnetizing value for a spherical shape).

\bibitem{Tholence} J.L. Tholence, {\it Solid State Comm.} {\bf 35},
113 (1980).

\bibitem{Mydosh} J.A. Mydosh, {\it Spin glasses: an experimental
introduction} (Taylor and Francis Ltd., London, 1993).

\bibitem{FerroUppsChaos}K. Jonason, J. Mattsson and P. Nordblad, {Phys. Rev. Lett.} {\bf
77}, 2562 (1996); R. Mathieu, P. Nordblad, D. N. H. Nam, N. X. Phuc and N. V. Khiem, {\it Phys.
Rev. B} {\bf 63}, 174405 (2001).

\bibitem{Balents} L. Balents, J.-P. Bouchaud and M. M\'ezard, {\it
J. Phys. I France} {\bf 6}, 1007 (1996); J.-P. Bouchaud in {\it Soft and Fragile Matter}, ed. M. E.
Cates and M. R. Evans (Institute of Physics Publishing, Bristol and Philadelphia, 2000) (preprint
cond-mat/9910387).


\bibitem{Levelut} J.-Ph. Bouchaud, P. Doussineau, T. de Lacerda-Ar\^oso and A. Levelut, {\it Eur. Phys. J. B} {\bf 21}, 335 (2001).

\bibitem{Berthier}  Rejuvenation and memory effects in the absence of static temperature
chaos have recently been found in simulations
 of the 4d Edwards Anderson model, see L. Berthier and J.-P. Bouchaud, preprint  condmat/0202069.

\bibitem{Martin} J. Houdayer and O.C. Martin, Europhys. Lett., {\bf
49}, 794 (2000);  J. Houdayer, F. Krzakala, O. C. Martin, {\it Eur. Phys. J. B}  {\bf 18}, 467-477
(2000).

\end{thebibliography}
\end{document}